\documentclass{ws-ijbc-arxiv}

\usepackage{ws-rotating}     % used only when sideways tables/figures are used
\usepackage{lscape}

\newcommand{\sz}{\scriptsize}

\begin{document}

\catchline{}{}{}{}{} % Publisher's Area please ignore

\markboth{Clara Granell {\it et al.}}{Paper Title}

\title{HIERARCHICAL MULTIRESOLUTION METHOD TO OVERCOME THE RESOLUTION LIMIT IN COMPLEX NETWORKS}

\author{CLARA GRANELL}
\address{Department d'Enginyeria Inform\`{a}tica i Matem\`{a}tiques, Universitat Rovira i Virgili\\
Av.\ Pa\"{\i}sos Catalans 26, 43007 Tarragona, Catalonia, Spain\\
clara.granell@urv.cat}

\author{SERGIO G\'OMEZ}
\address{Department d'Enginyeria Inform\`{a}tica i Matem\`{a}tiques, Universitat Rovira i Virgili\\
Av.\ Pa\"{\i}sos Catalans 26, 43007 Tarragona, Catalonia, Spain\\
sergio.gomez@urv.cat}

\author{ALEX ARENAS}
\address{Department d'Enginyeria Inform\`{a}tica i Matem\`{a}tiques, Universitat Rovira i Virgili\\
Av.\ Pa\"{\i}sos Catalans 26, 43007 Tarragona, Catalonia, Spain\\
alexandre.arenas@urv.cat}

\maketitle

\begin{history}
\received{(to be inserted by publisher)}
\end{history}

\begin{abstract}
The analysis of the modular structure of networks is a major challenge in complex networks theory. The validity of the modular structure obtained is essential to confront the problem of the topology-functionality relationship. Recently, several authors have worked on the limit of resolution that different community detection algorithms have, making impossible the detection of natural modules when very different topological scales coexist in the network. Existing multiresolution methods are not the panacea for solving the problem in extreme situations, and also fail. Here, we present a new hierarchical multiresolution scheme that works even when the network decomposition is very close to the resolution limit. The idea is to split the multiresolution method for optimal subgraphs of the network, focusing the analysis on each part independently. We also propose a new algorithm to speed up the computational cost of screening the mesoscale looking for the resolution parameter that best splits every subgraph. The hierarchical algorithm is able to solve a difficult benchmark proposed in \cite{lanci}, encouraging the further analysis of hierarchical methods based on the modularity quality function.
\end{abstract}


\section{Introduction}

\noindent
The quality function called {\em modularity} has been largely used in the assessment of the modular structure of networks \cite{firstnewman,newgirvan,newanaly,clauset,duch,jstat} and for data clustering and exploration \cite{newspect,granellchaos11}. Modularity is a global descriptor of a complex network that measures the difference between a given partition of the network and the same partition in an ensemble of the randomized versions of the original network preserving the local strength of every node. The optimization of modularity is coherently related to the definition of modules in the network; a module is defined as the  result of the optimal modularity partition. In 2007, Fortunato \& Barth\'elemy \cite{fortunato} pointed out a drawback in this function consisting in a certain resolution limit (generalized later in \cite{kumpula,good}), beyond which optimization of modularity is unable to identify certain modules, even those easily detectable at first sight, such as cliques almost disconnected from the rest of the network. This effect is known as the resolution limit of modularity. This problem arises because modularity fixes a global scale that could be appropriate for some networks but not for others, specially not suitable for those networks conformed by coexisting densely large and small communities. After this work, a multiresolution method was introduced in \cite{njp08}, which preserved the use of modularity with the addition of a parameter to control the resistance of nodes to form communities. The idea is that the analysis of communities may be performed at different scales of description, and the resolution limit is overcome just by moving to the right scale. Other  methods to overcome the resolution limit are found in \cite{rb,Pons,traagCPM,berry,peter,peter2}.

A recent work by \cite{lanci} shows that even those methods devoted to avoid the resolution limit, indeed have a resolution limit, and propose the use of an algorithm composed of several approaches called OSLOM \cite{lancichinetti} to really avoid such resolution problem. The proof that multiresolution schemes still have a resolution limit is performed analytically on the RB (after Reichardt-Bornholdt) method, and extended qualitatively using examples to the AFG (after Arenas-Fern\'andez-G\'omez) method and the recent CPM (Constant Potts Model) method.

\begin{figure}
  \begin{center}
    \mbox{\includegraphics*[width=.450\textwidth]{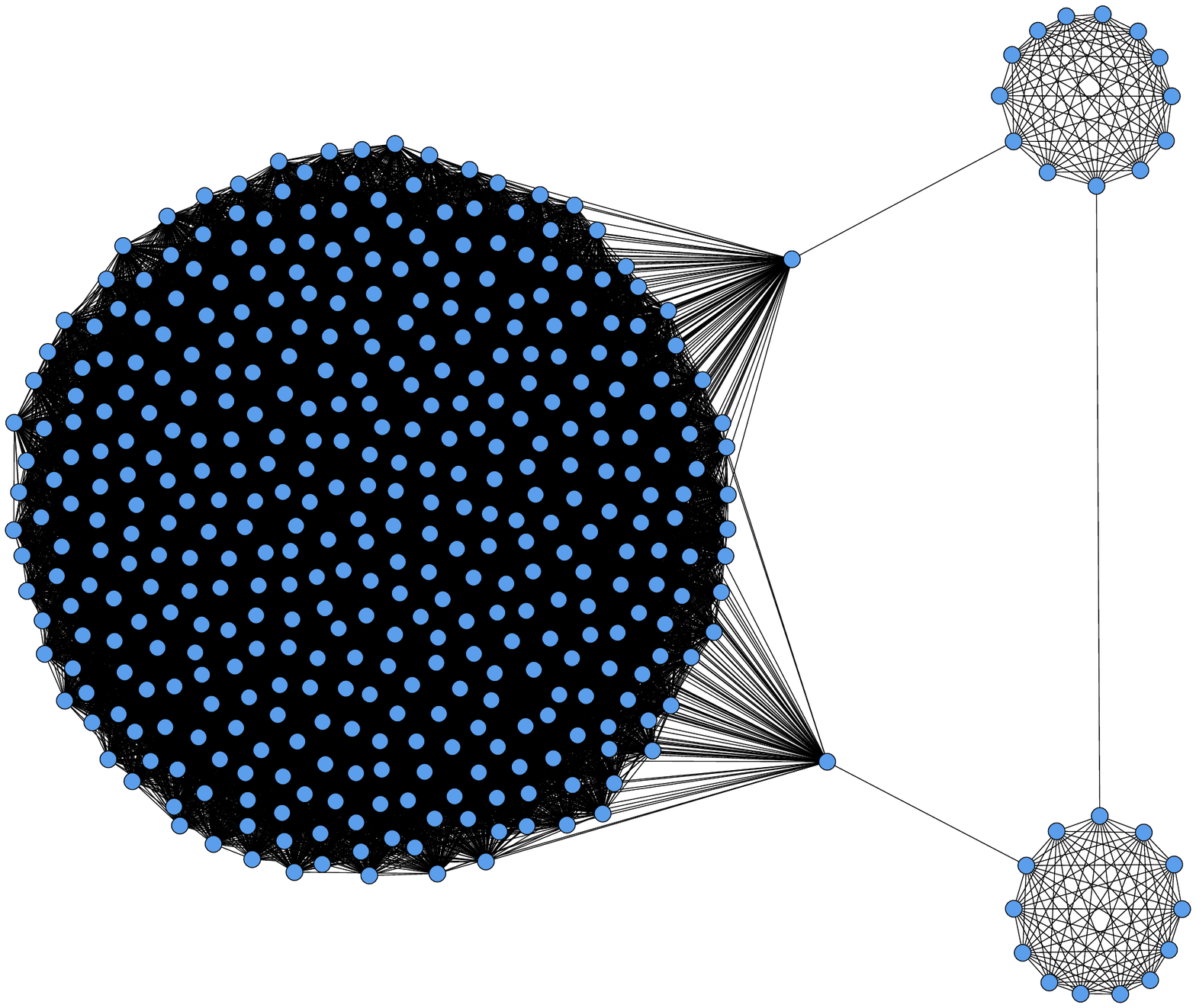}}
  \end{center}
  \caption{Benchmark proposed by \cite{lanci} to test the resolution limit of multiresolution methods. The large component is a ER network of 400 nodes with $k=100$ linked to two cliques of 13 nodes each, sharing only one link between them. The goal is to separate these three subgraphs using a community detection algorithm aimed to detect multiple resolutions.}
  \label{bench_fortu}
\end{figure}

We have performed extensive simulations using the AFG method and conclude that the authors of \cite{lanci} are right, the AFG method also has a resolution limit, and that the benchmark they propose (see Fig.~\ref{bench_fortu}), consisting of a giant Erd\"os-R\'enyi (ER) network and two small cliques, connected between them by just one link, is impossible to separate in the configuration of one cluster for the giant ER network and one cluster for each of the cliques, in the current proposal of the AFG method. Even though the synthetic benchmarks where multiresolution methods could fail are far from the structure of real networks, it is still challenging to investigate what are the problems and how to solve them.

In this paper we focus in the AFG method, analyzing its performance in resolution limiting situations, and proposing alternatives to eliminate or, if not possible, diminish the effect of this limit to minimum. An alternative is presented, a hierarchical application of the resolution screening, that avoids the resolution limit. The hierarchical application of a multiresolution method consists in to focus the screening on different clusters of the network as soon as these clusters are detected.
%This approach resembles the mechanism by which different parts of the brain integrate information \cite{meunier,jgg} using a hierarchical organization of its topology.

\section{Multiresolution AFG method}

In a previous work, the authors proposed a method that allows the full screening of the topological structure at different resolution levels using the original formulation and semantics of modularity as defined in \cite{firstnewman}. The original modularity allows the comparison of different partitions of the network. Given a network partitioned into communities, being $C_i$ the community to which node $i$ is assigned, the mathematical definition of modularity is
\begin{equation}
  Q[w_{ij},C]=\frac{1}{2w}\sum_i \sum_j \left(w_{ij}-\frac{w_i w_j}{2w}\right)\delta(C_i,C_j)\,,
  \label{QW}
\end{equation}
where $w_{ij}$ is the weight of the link between nodes $i$ and $j$ (zero if no link exists), $w_i=\sum_j w_{ij}$ is the strength of node $i$  and  $2w=\sum_i w_i$ is the total strength of the network \cite{newanaly}. The Kronecker delta function $\delta(C_i,C_j)$ takes the value 1 if node $i$ and $j$ are into the same community and 0 otherwise. Several authors have attacked the problem of modularity optimization, with considerable success, by proposing different heuristics \cite{newfast,clauset,rogernat,duch,pujol,newspect}, see \cite{fortunato_rev} for a review.

The AFG method was designed to evaluate the community structure of networks using a kind of magnifying glass of the topology \cite{njp08}. The mathematical form of this prescription is given by
\begin{equation}
  Q_{\mbox{\sz AFG}}[w_{ij},C,r]=Q[w_{ij}+ r \delta_{ij},C]\,,
  \label{QWr}
\end{equation}
where the resistance $r$ is the parameter controlling the resolution of the partitions we want to find, and $w_{ij}+ r \delta_{ij}$ is the new weights matrix after the addition of a self-loop with value $r$ to each node. When $r$ is zero, we recover the standard modularity $Q$. The definition of $Q_{\mbox{\sz AFG}}$ preserves the original semantics of modularity.

A refinement of the AFG method may be found in \cite{granellchaos11,granellijbc12}, where the original formulation of modularity Eq.~(\ref{QW}) is replaced by its extension to networks with positive and negative weights \cite{signed,traag}. Although the differences are usually small, this is necessary since the access to the macroscale needs the use of negative values of the resistance, even if the original network has only positive weights. Thus, the adequate formulation of modularity Eq.~(\ref{QW}) for undirected weighted signed networks which should be used is
\begin{equation}
  Q[w_{ij},C] = \frac{1}{2w^{+} + 2w^{-}} \sum_i \sum_j
        \left[ w_{ij} - \left(
          \frac{w_i^{+} w_j^{+}}{2w^{+}} - \frac{w_i^{-} w_j^{-}}{2w^{-}}
        \right) \right]
        \delta(C_i,C_j)\,.
  \label{QWS}
\end{equation}
where
\begin{eqnarray}
  w_i^{+} & = & \sum_{j,w_{ij}>0} w_{ij}\,,
  \\
  w_i^{-} & = & \sum_{j,w_{ij}<0} |w_{ij}|\,,
\end{eqnarray}
are the positive and negative strengths of node $i$, and
\begin{eqnarray}
  2w^{+} & = & \sum_i w_i^{+}\,,
  \\
  2w^{-} & = & \sum_i w_i^{-}\,,
\end{eqnarray}
are the positive and negative total strengths respectively. Please note that these four strengths are defined to be non-negative.
The extension to directed networks \cite{njp07} is simply obtained by the substitutions
in Eq.~(\ref{QWS})
\begin{eqnarray}
  w_{i}^{\pm} &\rightarrow &  w_{i}^{\pm,\mbox{\scriptsize out}} = \sum_{k} w^{\pm}_{ik}\,,\\
  w_{j}^{\pm} &\rightarrow &  w_{j}^{\pm,\mbox{\scriptsize in}} = \sum_{k} w^{\pm}_{kj} \,.
\end{eqnarray}
For the sake of simplicity, we will refer to the undirected case for the rest of the paper.
In the particular case that the original network does not have negative weights, $r<0$ and $w_{ii}=0$, $\forall i$, Eq.~(\ref{QWr}) reads
\begin{eqnarray}
  Q_{\mbox{\sz AFG}}[w_{ij},C,r] & = & \frac{1}{2w + N |r|} \sum_i \sum_j
        \left[ w_{ij} + r\delta_{ij} - \left(
          \frac{w_i w_j}{2w} - \frac{r^2}{N |r|}
        \right) \right]
        \delta(C_i,C_j)
  \\
  & = & \frac{2w}{2w - N r} Q[w_{ij},C]
        + \frac{r}{2w - N r} \left(N - \frac{1}{N}\sum_{s\in C} n_s^2\right)\,,
  \label{QWSr}
\end{eqnarray}
where $N$ is the total number of nodes, $n_s$ is the number of nodes in community $s$ of the partition $C$, and the nodes and total strengths refer to the original network before the addition of the self-loops. It is interesting to realize that, since all the negative strengths are equal to the absolute value of $r$, the contribution of the resistance to modularity is equivalent to a constant Potts model \cite{traagCPM}.

\begin{figure}
  \begin{center}
    \mbox{\includegraphics*[width=.750\textwidth]{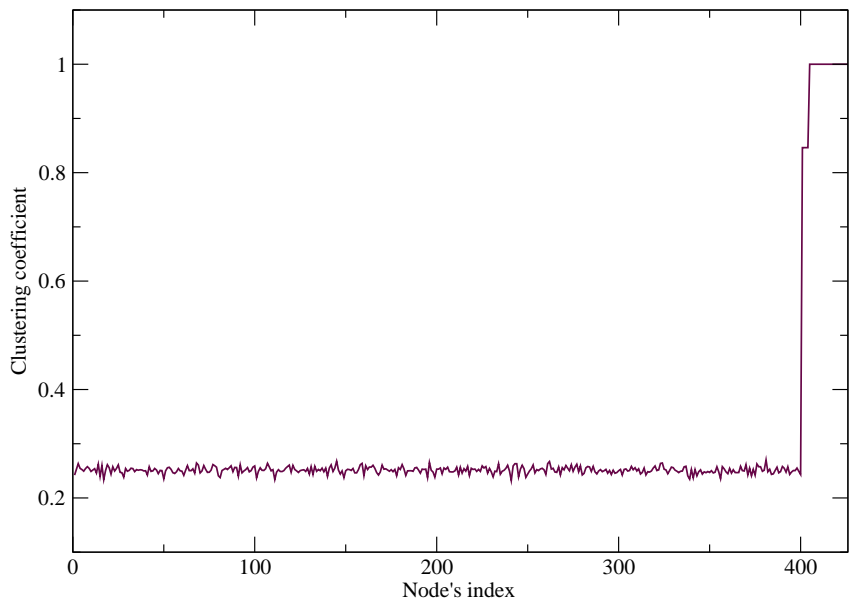}}
  \end{center}
  \caption{Clustering coefficient for the benchmark of Fig.\ref{bench_fortu}. Note the sharp transition in the relative local density of links represented by the clustering coefficient. This is indeed a hint to the coexistence of very different topological scales in the network.}
  \label{clust}
\end{figure}

Resolving the substructure of networks using a unique parameter as proposed in the AFG has still a resolution problem. As pointed out by \cite{lanci}, when very different sized modules coexist, multiresolution methods will tend to break the larger groups before finding the smaller ones. The phenomenon is easy to understand with an example: let us imagine an image with a real size elephant and an ant, to see the details of the ant we have to get so close to the image that the elephant image disintegrates in smaller parts, and only part of the elephant is seen when focusing on the ant. In terms of modularity, we are trying to unravel those areas which are denser in terms of links with respect to other areas in the network. A way to determine if we could have resolution problems is to plot a link density map and detect if there are sharp contrasts. If very different topological scales coexist, there will also be jumps in the clustering coefficient. In the example provided by \cite{lanci}, which consists of an ER network of 400 nodes with average degree 100 linked to two cliques of 13 nodes only by a unique link between them (see Fig.~\ref{bench_fortu}), the clustering coefficient presents a drastic separation of scales, see Fig.~\ref{clust}. This indicates small zones of the network very densely connected and a wide area not so dense, corresponding to the cliques and the ER, respectively.

\section{Hierarchical Multiresolution method}

Our approach to solve the resolution problem takes advantage of the capability of the AFG method to find meaningful communities from the initial steps of the mesoscale analysis. More precisely, we propose the use of an iterative scheme which combines the optimization of modularity close to the macroscale of the network with its splitting in subgraphs, one for each of the previously found communities.

\begin{figure}
  \begin{center}
    \mbox{\includegraphics*[width=.750\textwidth]{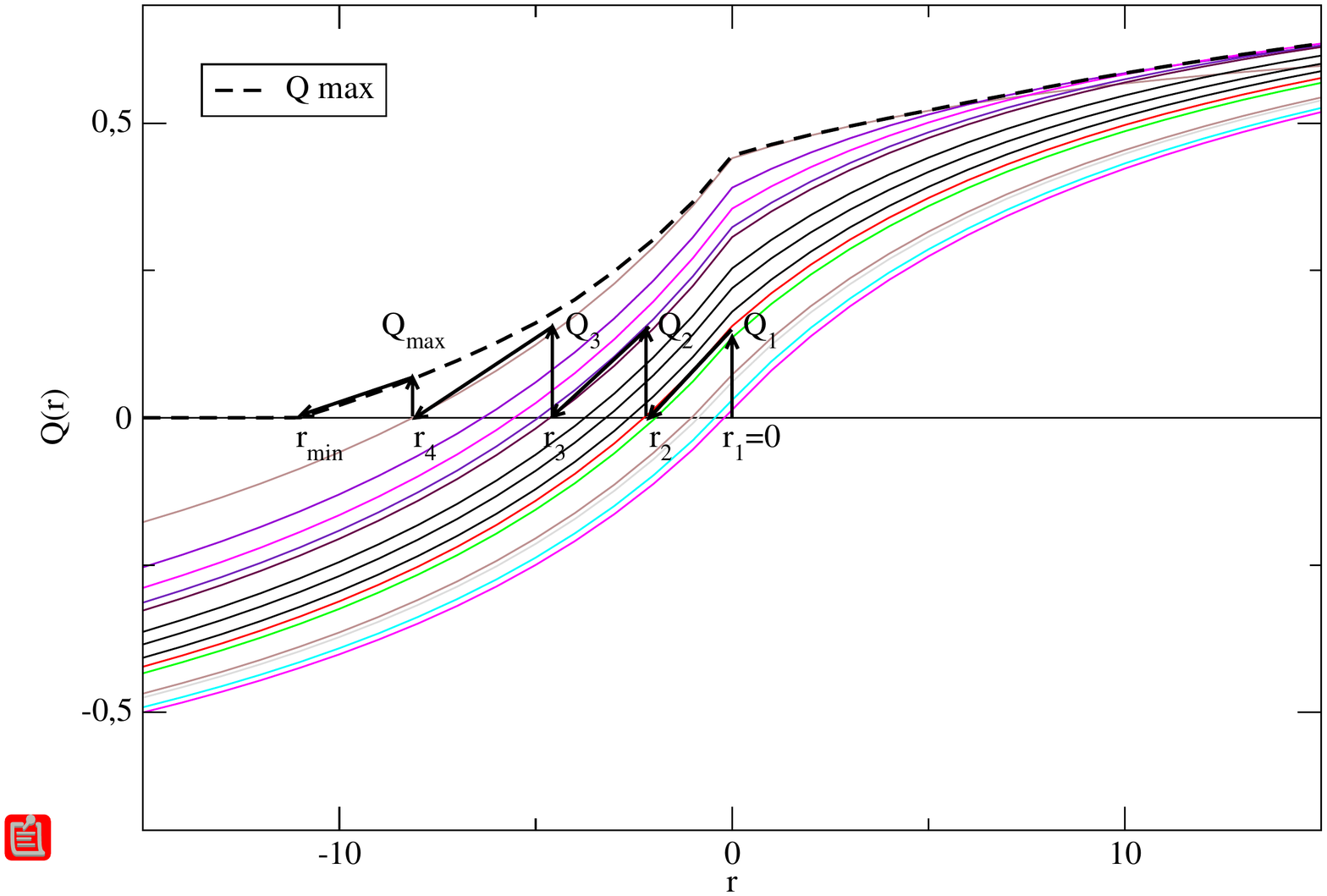}}
  \end{center}
  \caption{Example of evolution of the FTR algorithm finding $r_{\mathrm{min}}$ in four iterations of the scheme. We start at $r_1=0$, optimizing modularity we find $Q_1$, we look for the $r_{\mathrm{min}}$ corresponding to the partition found at $Q_1$ using Eq.~\ref{rminC} and label it $r_2$, the process follows with $Q_2 \rightarrow r_3 \rightarrow Q_3 \rightarrow r_4 \rightarrow Q_{\max}$ up to finding $r_{\mathrm{min}}$, beyond this value the only partition we will find corresponds to the whole network as a unique module. Different curves in color are values of $Q[w_{ij},{\cal C},r]$ for different partitions.}
  \label{ftr}
\end{figure}

Supposing that our network is undirected, weighted, with positive weights and no self-loops, the prescription of our algorithm is the following:
\begin{enumerate}
\item Start out from the macroscale partition ${\cal M}$, which has only one community containing all nodes. Then, find the upper bound of this macroscale, which is the minimum value of the resistance parameter ($r_{\min}$) needed to find a partition ${\cal C}$ of the network with optimal modularity $Q_{\mbox{\sz AFG}}[w_{ij},{\cal C},r_{\min}]$ formed by more than one community.
\item Split the network in the subgraphs defined by the partition ${\cal C}$ just found.
\item Repeat the previous steps with each subgraph until no further subdivisions are needed.
\end{enumerate}
This algorithm defines a hierarchical organization of the nodes, where the values of $r_{\min}$ at each splitting define the ultrametric distances between nodes, i.e.\ the heights in the dendrogram at which every pair of nodes first meet.

The calculations of $r_{\min}$ and ${\cal C}$ may be performed simultaneously, therefore avoiding the costly scanning of the whole mesoscale between the lower and upper bounds of the resistance \cite{granellchaos11,granellijbc12}. This is a consequence of the following properties:

\begin{itemlist}
\item The value of $r_{\min}$ is negative, with the only exception in which the network is just a clique.
\item $Q_{\mbox{\sz AFG}}[w_{ij},{\cal M},r] = 0$, $\forall r<0$, because:
\begin{eqnarray}
Q_{\mbox{\sz AFG}}[w_{ij},{\cal M},r] &=& \frac{1}{2w + N |r|} \sum_i \sum_j
        \left[ w_{ij} + r\delta_{ij} - \left(
          \frac{w_i w_j}{2w} - \frac{r^2}{N |r|}
        \right) \right] \nonumber\\
&=& \frac{1}{2w + N |r|}  \left[ 2w + N r - \left(\frac{(2w)^2}{2w} + \frac{N^2 r^2}{N r} \right) \right] = 0\,.
\end{eqnarray}
In fact, modularity Eq.~\ref{QWS} is always zero for ${\cal M}$, no matter the network or the value of the self-loops.
\item Since $Q_{\mbox{\sz AFG}}[w_{ij},{\cal M},r] = 0$ and modularity is a continuous and monotonically increasing function of the resistance for any given $C\neq {\cal M}$, the optimal partition ${\cal C}$ at $r_{\min}$ must satisfy $Q_{\mbox{\sz AFG}}[w_{ij},{\cal C},r_{\min}] = 0$.
\item For any given partition $C$, the minimum meaningful value of the resistance $r_{\min}(C)$ is the one for which $Q_{\mbox{\sz AFG}}[w_{ij},C,r_{\min}(C)] = 0$. Thus, Eq.~(\ref{QWSr}) leads to
  \begin{equation}
    r_{\min}(C) = \frac{-2w}{\ds N - \frac{1}{N}\sum_{s\in C} n_s^2}\, Q[w_{ij},C]\,.
    \label{rminC}
  \end{equation}
\item The upper bound of the macroscale is given by
  \begin{equation}
    r_{\min} = \min_{C}\left\{ r_{\min}(C) \right\}\,
    \label{rmin}
  \end{equation}
  and ${\cal C}$ is the partition which minimizes $r_{\min}(C)$.
\end{itemlist}

All these properties may be combined in the following {\em fast-tracking resistance} (FTR) algorithm to find the upper bound of the macroscale:
\begin{enumerate}
\item Optimize modularity at $r=0$, to obtain partition $C_{\mbox{\sz prev}}$.
\item Calculate $r_{\min}(C_{\mbox{\sz prev}})$ using Eq.~(\ref{rminC}).
\item Optimize modularity at $r:=r_{\min}(C_{\mbox{\sz prev}})$, to obtain the current partition $C_{\mbox{\sz curr}}$.
\item If $C_{\mbox{\sz curr}}=C_{\mbox{\sz prev}}$ or $C_{\mbox{\sz curr}}={\cal M}$, then $r_{\min}:=r_{\min}(C_{\mbox{\sz prev}})$ and ${\cal C}:=C_{\mbox{\sz prev}}$.
\item Otherwise, let $C_{\mbox{\sz prev}}:=C_{\mbox{\sz curr}}$ and go back to the second step.
\end{enumerate}
In practice, this algorithm converges in a few number of steps. It stops when a value of $r$ is found such that the optimization of modularity does not produce any new partition. In this case, the modularity of both $C_{\mbox{\sz prev}}$ and ${\cal M}$ is zero, and no known partition can be used to obtain a better upper bound of the macroscale. Of course, we cannot claim that we have found the ``real'' $r_{\min}$, since no optimization heuristic can ensure the finding of the global maximum of modularity (this problem is known to be a NP-hard problem, see \cite{brandes}), but this is the best approximation one may obtain. To exemplify the functioning of the FTR algorithm we show in Fig.~\ref{ftr} its application to the first hierarchical splitting of Zachary karate club network \cite{zach}.

\section{Results and discussion}

We have applied the hierarchical multi-resolution method explained before to the benchmark proposed by \cite{lanci} shown in Fig.~\ref{bench_fortu}. We use the FTR algorithm to speed up the process of finding the minimal $r$ at which every subgraphs splits.
The aim is to find the partition divided in three communities in which the giant ER and each clique are separated. These three communities should contain the nodes labeled 1 to 400, 401 to 413 and 414 to 426, respectively.
\begin{figure}[t]
\begin{center}
    \mbox{\includegraphics*[width=.550\textwidth]{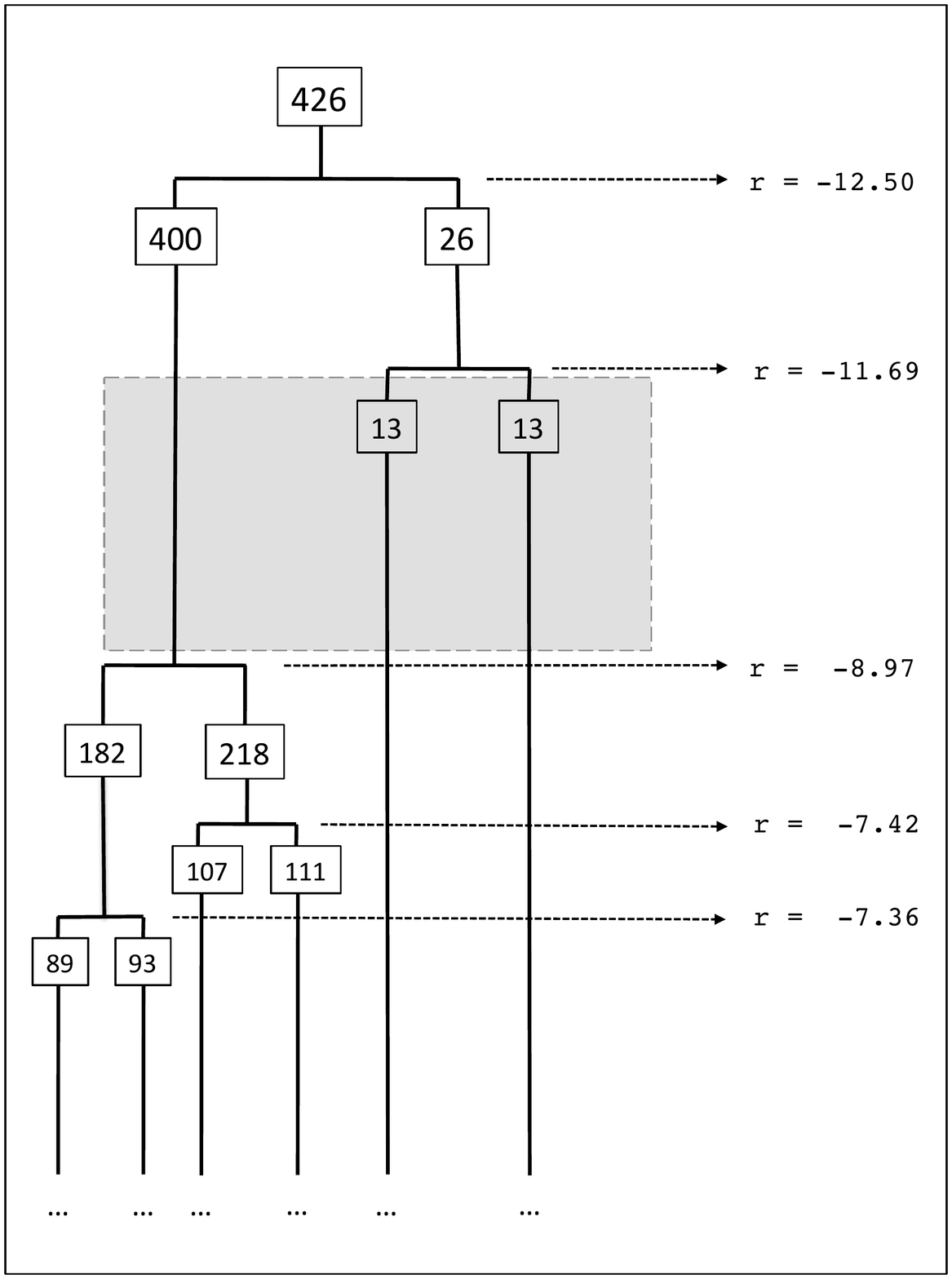}}
  \end{center}
  \caption{Dendrogram resulting from the application of the hierarchical multiresolution method on the benchmark of Fig.~\ref{bench_fortu}. The grey region shows the range of the resistance parameter in which the three communities searched coexist. Note that the vertical lines are not scaled.}
  \label{dendo}
\end{figure}
As stated in the method, we have started out from the macroscale ${\cal M}$ of the network, which contains the 426 nodes. The optimal partition splits in two communities at a value of the resistance parameter -12.5, obtaining a community formed by the nodes from 1 to 400 and another community containing the 26 nodes corresponding to the two cliques. Performing the hierarchical method on the two communities obtained, we find that the community containing the 26 nodes rapidly splits in two communities of 13 nodes, at a value of the resistance equal to -11.69. The partition containing 400 nodes splits in two at a much greater value of the resistance parameter, which is -8.97. After that, a hierarchical multiresolution is applied to any community found, until no further divisions are needed. The results of this example are shown in a dendrogram representation in Fig.~\ref{dendo}.

Observing this figure, we find that there is a region of the resistance parameter in which the three communities we were hoping to find coexist. This happens because the two cliques form their own communities much before the community of 400 nodes is split in two. Note that this result can not be obtained using the original multiresolution AFG method exploring the whole mesoscale, because of the resolution limit emerging from the coexistence of very different topological scales. The rationale behind the success of the hierarchical method in this situation is the following: the separation of the network in optimal subgraphs, each one split and independently analyzed through the multiresolution scheme, reduces the global resolution limit. This resolution limit depends on the number of nodes and the number of links in the whole structure. The multiresolution method is able to focus the attention on lower scales while other parts of the network are being screened independently at larger resolution values of $r$.

\section{Conclusions}

We have presented a hierarchical multiresolution method able to cope with networks where the resolution limit would make other schemes to fail, finding the natural communities as defined by \cite{fortunato}. The method is boosted by a mechanism that allows the determination of the resolution parameter $r$ at which to optimize modularity in a few steps. The results solving the difficult separation of the benchmark proposed in \cite{lanci} are encouraging and open the door for further investigation of modularity based community detection methods to escape from the implicit resolution limit.

\section*{Acknowledgements}
We acknowledge support from the Spanish Ministry of Science and Innovation FIS2009-13730-C02-02
and the Generalitat de Catalunya SGR-00838-2009.
\bibliographystyle{ws-ijbc}
\bibliography{cga_ijbc_rev}

\end{document}